\documentclass[aps,preprint,showpacs,a4paper,nofootinbib]{revtex4}
\usepackage{bm}
\usepackage{graphicx}  
\begin{document}

\title{On the heavy Majorana neutrino and light sneutrino contribution
to \bm{$e^{-}e^{-}\;\to\;\ell^{-} \ell^{-}$}, (\bm{$\ell=\mu,\tau)$} }

\author{{M.~Cannoni},$^{1,2}$ {St.~Kolb},$^{2}$ and {O.~Panella}$^{2}$}

\affiliation{
$^{1}$Dipartimento di Fisica, Universit\`a degli Studi di Perugia,
Via A. Pascoli, I-06123, Perugia, Italy
}
\affiliation{
$^{2}$Istituto Nazionale di Fisica Nucleare, Sezione di Perugia,
 Via A.~Pascoli, I-06123 Perugia, Italy
}

\email[Corresponding author: ]{mirco.cannoni@pg.infn.it}

\date{February 25, 2003}

\begin{abstract}
The cross section for the  reaction $e^{-}e^{-}\;\to\; \ell^-\ell^-$ 
($\ell=\mu,\tau$) 
is calculated in models with heavy Majorana neutrinos mediating lepton 
number violating amplitudes at the loop level. The contributing
four-point functions are evaluated exactly (numerically) taking into account 
the full propagator dependence on external momenta, thereby extending 
to the energy range of interest for the next linear colliders 
an earlier approximate low energy calculation. 
The amplitude shows a non-decoupling behaviour relative to the 
heavy Majorana neutrino masses, but due to the stringent 
bounds on heavy-light mixing the signal cross section attains  
observable values only for the less constrained $\tau$ signal.
The cross section induced by lepton number violation in the
$SU(2)_L$ doublet sneutrino sector of supersymmetric extensions of
the standard model is constrained by the upper limits on neutrino masses 
and probably too tiny to be observable.

\end{abstract}

\pacs{12.60.Cn, 12.60.Jv, 13.15.+g, 14.60.St }

\maketitle

\begin{figure}
\scalebox{0.5}{\includegraphics*[75,225][510,730]{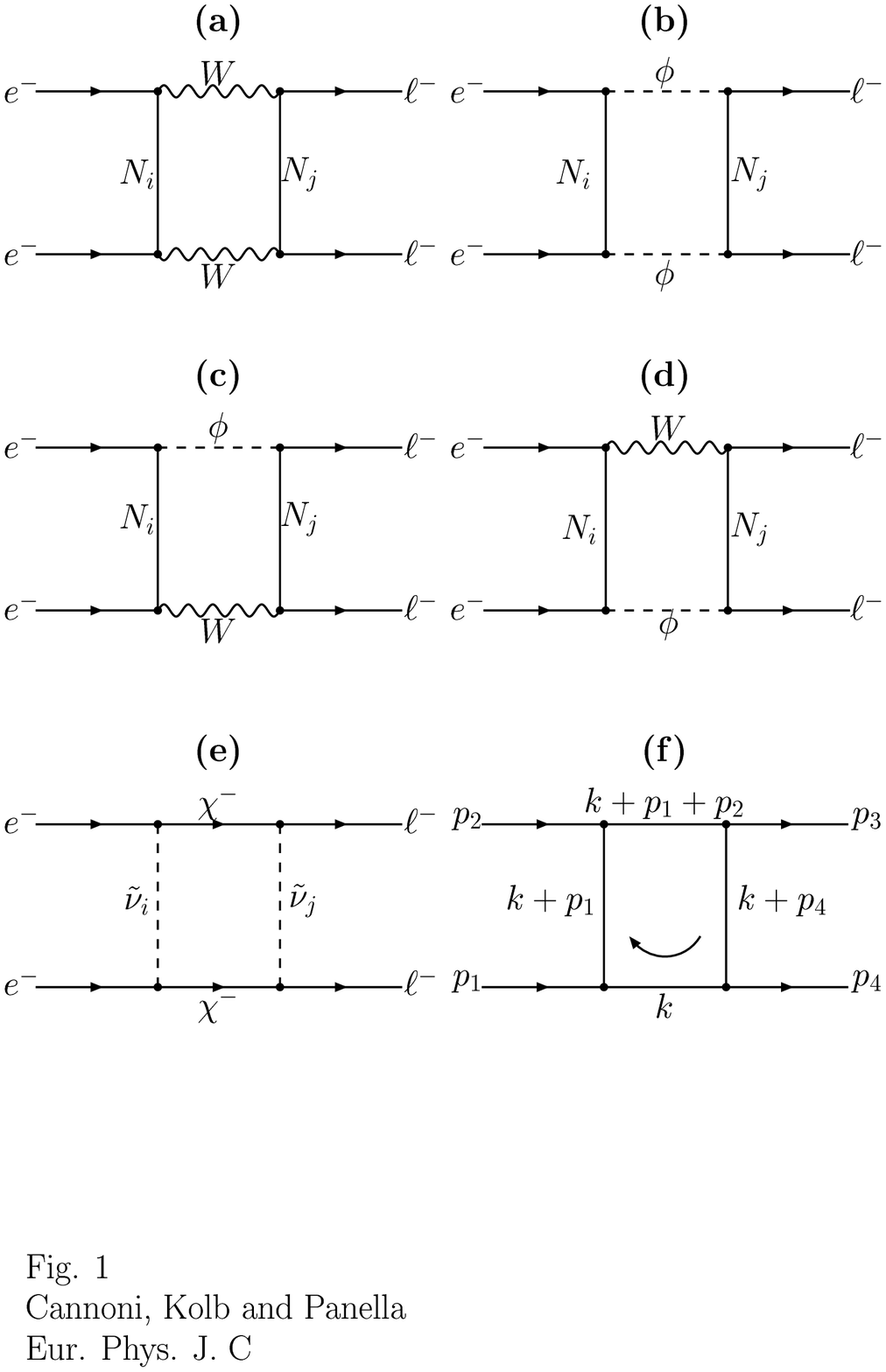}}%
\caption{In (a-d) the Feynman diagrams, in the 't Hooft-Feynman 
gauge, contributing to $e^-e^- \to \ell^-\ell^- $ ($\ell=\mu,\tau$) via heavy Majorana neutrinos. 
In (e) the corresponding SUSY diagram is given which arises 
in models with lepton number violation in the sneutrino sector. 
In (f) the choice of the running momentum is given. 
The momenta $k_i$ ($i=1,2,3$) for the decomposition of the tensor 
integrals  within the  {\scshape looptools} notation are:
 $k_{1}=p_{1}$, $k_{2}=p_{1}+p_{2}$, $k_{3}=p_{1}+p_{2}-p_{3}=p_{4}$.}
\label{fig1}
\end{figure}

The process $e^{-}e^{-}\;\to\;\ell^{-}\ell^{-}$ ($\ell=\mu,\tau$),
which violates the $L_e$ and $L_{\ell}$ lepton  numbers, 
is  forbidden in the Standard Model (SM) due 
to exact lepton number  conservation to all orders of perturbation theory. 
Its observation at a next 
generation linear collider with center of mass energy $\sqrt{s}=500,\;800,
\;1000$ GeV may be possible only if there is new physics  
that can trigger it.
The signature is clear and practically free from SM background.
In literature it was studied: $(i)$ in the context of models with gauge 
bileptons~\cite{Frampton:2000sd}, where the final state is reached 
through tree level s-channel annihilation into a gauge bilepton 
and subsequent decay,
$(ii)$ in the context of mixing models  where the reaction proceeds 
through a loop (box diagram) with heavy Majorana 
neutrinos and W$^{-}$ gauge bosons as virtual particles running in the 
loop~\cite{Pham:2000bz}, and $(iii)$ in supersymmetric scenarios where
sneutrinos and charginos instead of neutrinos and charged bosons are 
exchanged~\cite{halprin}. 
The aim of this paper is: $(i)$ to improve and extend the calculation 
of Ref.~\cite{Pham:2000bz} (which was essentially a low energy calculation)
in order to provide predictions in the energy range
of interest for the next linear collider project, motivated by 
the observation that all diagrams of the 
process, see Fig.~\ref{fig1}(a-d), have a threshold singularity
at $\sqrt{s} = 2 M_W$ where the amplitude develops an imaginary part
giving a boost to its absolute value 
(see the well known example of photon-photon 
scattering~\cite{landau}), taking also into account experimental 
bounds on effective mixing angles not considered in ~\cite{Pham:2000bz};
$(ii)$ to make a realistic calculation for the sneutrino case: the cross
section was estimated so far assuming eV scale sneutrinos~\cite{halprin}.

\begin{figure}
\begin{center}
\scalebox{0.6}{\includegraphics*[30,295][570,702]{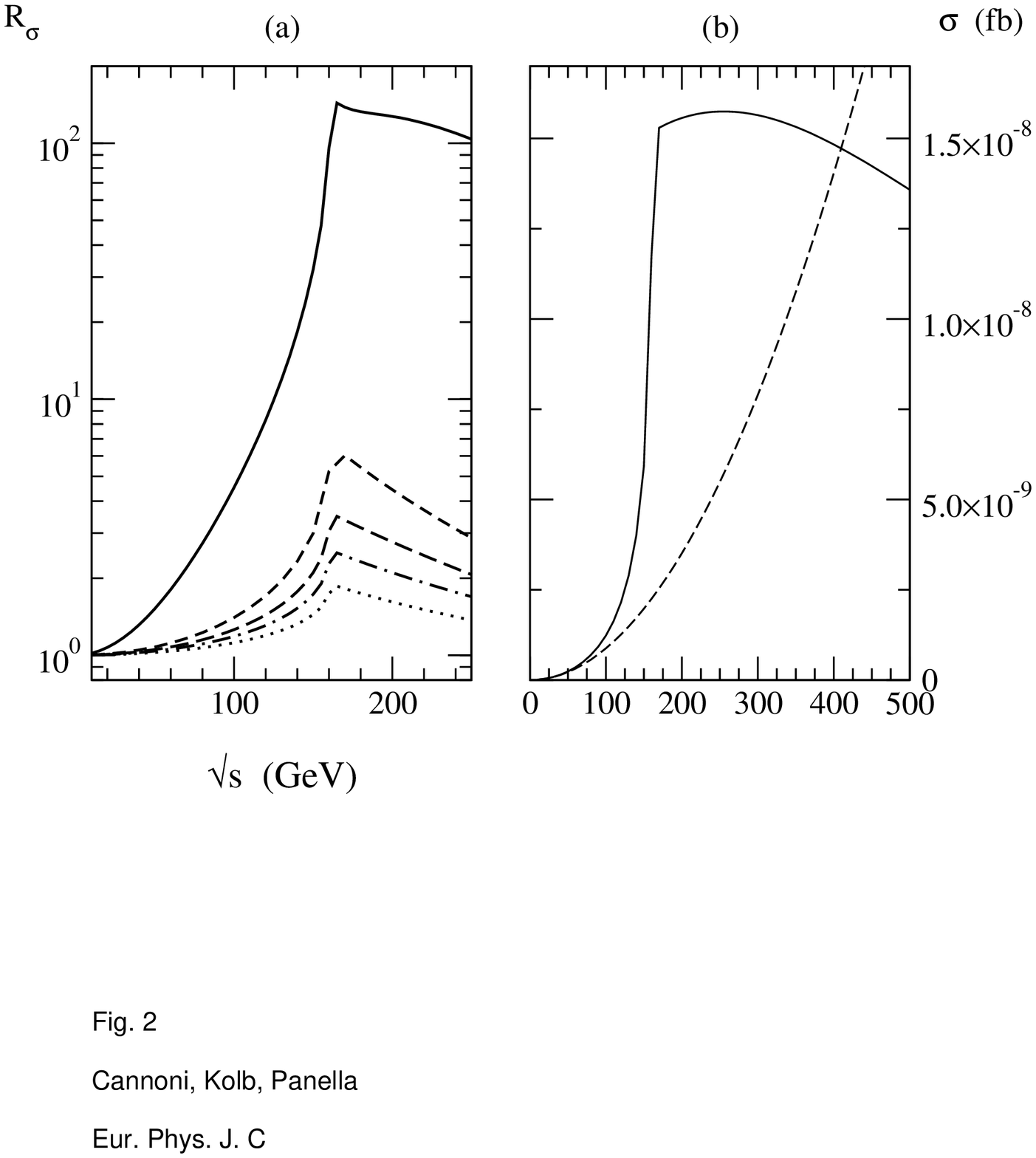}}
\caption{
In (a) the ratio $R_\sigma$ is plotted as a function of $\sqrt{s}$ the 
energy in the center of mass system: solid line, 
$M_{N_i}=M_{N_j}=100$ GeV; short-dashed line, $M_{N_i}=150$ GeV, $M_{N_j}=450$
GeV; long-dashed line, $M_{N_i}=M_{N_j}=500$ GeV;
dot-dashed line, $M_{N_i}=M_{N_j}=1$ TeV; dotted line, $M_{N_i}=M_{N_j}=3$ TeV.
In (b) the absolute values of the cross sections are given for a particular choice 
of Majorana masses $M_{N_i}=150$ GeV, $M_{N_j}=450$ 
as function of the energy in the  
center of mass frame.  The solid line is obtained integrating 
Eq.~(\protect\ref{cross})
while the dashed line is based on Eq.~(9) of~\cite{Pham:2000bz}.
}
\label{fig2}
\end{center}
\end{figure}

We assume that heavy Majorana neutrinos (mass eigenstates) 
couple to the standard model charged currents through heavy-light neutrino mixing. 
This is  the simplest way to obtain lepton number violating processes.
Consider the box diagrams depicted in Fig.~\ref{fig1}(a-d). 
As in~\cite{Pham:2000bz}, we use the 't Hooft-Feynman
gauge: there are graphs with $WW$, $\phi\phi$ and
 $\phi{W}$ exchange, $\phi$ being the Goldstone boson.
The lagrangian of interest is~\cite{Ilakovac:1994kj}:
\begin{eqnarray}
{\cal L}&=&\sum_{\ell,N_{i}}
-i\frac{g}{\sqrt{2}}\left[{\bar{\psi}}_{\ell}\gamma_\mu\frac{1-\gamma
_{5}}{2}U_{\ell N_{i}} \psi_{N_{i}} W^{\mu} \right.\cr 
&-&\left. \frac{M_{N_{i}}}{M_{W}}{\bar{\psi}}_{\ell}\frac{1+\gamma_{5}}{2}
U_{\ell N_{i}}\psi_{N_{i}} \phi\right]+h.c.
\label{lag}
\end{eqnarray}
where $\ell=e,\mu,\tau$ and $U_{\ell N_{i}}$ are the elements of the 
mixing matrix of the heavy mass eigenstates $M_{N_{i}}$ labeled 
by the index $N_{i}$. Neglecting the masses of the external particles 
allows to simplify the calculation of the amplitudes that 
can be expressed in terms of: $(i)$ the Mandelstam variables $s$, $t$ 
and $u$; $(ii)$ the spinor products of light-like momenta 
(see Ref.~\cite{Panella:2001wq} and references therein)
\begin{eqnarray}
S(p_a,p_b)&=&\bar{u}_+(p_a) \cdot u_-(p_b)\cr
T(p_a,p_b)&=&\bar{u}_-(p_a) \cdot u_+(p_b),
\end{eqnarray}
which obey the relation: $|S(p_2,p_1)\, T(p_3,p_4) |^2 = s^2$; 
and $(iii)$ the scalar $D_0$ and the tensor  rank-1 ($D_\mu$) and rank-2  
($D_{\mu\nu}$) four-point functions~\cite{tooft}.
The corresponding amplitudes are found to be:
\begin{eqnarray}
{\cal M}^{WW}_{a}&=&\left(\frac{g}{\sqrt{2}}\right)^{4}
\frac{1}{(4\pi)^{2}}\sum_{N_{i},N_{j}}
(U_{eN_{i}}^{*}U_{\mu N_{j}})^{2} M_{N_{i}} M_{N_{j}}\cr
&\times& 4 S(p_2,p_1)T(p_3,p_4)\left[D_{0}(s,t)+D_{0}(s,u)\right]\, ,
\label{ampa} 
\end{eqnarray}
\begin{eqnarray}
{\cal M}^{\phi\phi}_{b}&=&\left(\frac{g}{\sqrt{2}}\right)^{4}
\frac{1}{(4\pi)^{2}}\sum_{N_{i},N_{j}}
(U_{eN_{i}}^{*}U_{\mu N_{j}})^{2}
\frac{M^{2}_{N_{i}}}{M^{2}_W}\frac{M^{2}_{N_{j}}}{M^{2}_W}\cr
&\times& M_{N_{i}}M_{N_{j}}
S(p_2,p_1)T(p_3,p_4)\left[D_{0}(s,t)+D_{0}(s,u)\right]\, ,
\label{ampb} 
\end{eqnarray}
\begin{eqnarray}
{\cal M}^{W \phi}_{c}&=&\left(\frac{g}{\sqrt{2}}\right)^{4}
\frac{1}{(4\pi)^{2}}\sum_{N_{i},N_{j}}(U_{eN_{i}}^{*}U_{\mu N_{j}})^{2}
\frac{M_{N_{i}}}{M_W}\frac{M_{N_{j}}}{M_W}\cr
&\times& S(p_2,p_1)T(p_3,p_4) \left[ 4(D_{00}(s,t)+D_{00}(s,u))\right.\cr
&-& \left. 2 (tG(s,t)+uG(s,u)) \right.\cr
&-& \left. 2 {(tV(s,t)+uV(s,u))}\right]\, ,
\label{ampc}
\end{eqnarray}
\begin{eqnarray}
{\cal M}^{\phi W}_{d}&=&\left(\frac{g}{\sqrt{2}}\right)^{4}
\frac{1}{(4\pi)^{2}}\sum_{N_{i},N_{j}} (U_{eN_{i}}^{*}U_{\mu N_{j}})^{2}
\frac{M_{N_{i}}}{M_W}\frac{M_{N_j}}{M_W}\cr
&\times& S(p_2,p_1)T(p_3,p_4)\left[ 4(D_{00}(s,t)+D_{00}(s,u))\right.\cr
&-& \left.2 (tG(s,t)+uG(s,u))\right]\, .
\label{ampd} 
\end{eqnarray}
The numerical computation of the four-point functions was performed 
using the 
{\scshape{looptools}} \cite{Hahn:1998yk} software where the following notation 
is used as regards the expansion of the rank-1 and rank-2 
tensor functions: $D_{\mu}=\sum_{i=1}^{3} D_{i}\, ({k_i})_\mu$, 
$D_{\mu\nu}=g_{\mu\nu}D_{00} + 
\sum_{i,j=1}^{3}\,D_{ij}\, ({k_i})_\mu \, ({k_j})_\nu$, 
$k_{i}$ being sums of external momenta running in the loop as 
explained in Fig.~\ref{fig1}(f). Within this notation the form factors 
$G$ and $V$ appearing in Eqs. 
(\ref{ampa})-(\ref{ampd}) are given by 
\begin{eqnarray}
G&=&D_{22}+D_{23}+D_{12}+D_{13}\, ,\cr
V&=&2D_{2}+D_{1}+D_{3}+D_{0}\,.
\nonumber
\end{eqnarray}
Terms depending both on  $(s,t)$ and $(s,u)$ appear because the  
identical fermions in the final state require proper 
anti-symmetrization of the amplitudes.
Defining $x_W=\sin^{2}{\theta_W}$ and 
$x_{i,j}={M^{2}_{N_{i,j}}}/{M^{2}_W}$,
the differential cross section is easily found to be:
\begin{eqnarray}
\frac{d\sigma}{d\cos{\theta}}=\frac{1}{256\pi}
\left(\frac{\alpha}{x_W}\right)^{4}|K(s,t,u)|^{2}s,
\label{cross}
\end{eqnarray}
where $K$ is given by:
\begin{eqnarray}
K&=&\sum_{N_{i},N_{j}} (U_{eN_{i}}^{*}U_{\mu N_{j}})^{2}
\sqrt{x_{i}x_{j}}\left\{M_{W}^{2}\left(1+\frac{x_{i}x_{i}}{4}\right)\right.\cr
&\times& \left.\left[D_{0}(s,t)+D_{0}(s,u)\right]+2\left(D_{00}(s,t)+D_{00}(s,u)
\right)
\right.\cr
&-& \left. [tG(s,t)+uG(s,u)]-\frac{[tV(s,t)+uV(s,u)]}{2} \right\}.
\label{result}
\end{eqnarray}
To obtain the total signal cross section $\sigma_{tot}$, 
Eq.~(\ref{cross}) is integrated 
numerically over the scattering angle in the center of mass frame.
As stated above, similar formulas were obtained 
in~\cite{Pham:2000bz} using the approximation where {\em all external 
momenta in the loops are  neglected} relative to the heavy masses 
of the gauge bosons and Majorana neutrinos, enabling to carry 
out the loop integration analytically.  
The formulas thus obtained are well  known in the 
literature~\cite{Inami:1980fz} and   the final cross section, which 
depends only on $x_{i,j}$ and the mixing coefficients, grows 
linearly with  the center of mass energy squared, $s$.
This approximation for the four-point functions is good at low energies, 
such as in decay processes of heavy mesons,
or when $\sqrt{s}<<M$, $M$ being the highest mass running in the loop. 
In addition the linear 
growth with $s$ would break unitarity, therefore in order
to make quantitative predictions with the correct high energy behaviour, 
the four-point functions  
full dependence on the external momenta has to be considered. 
Theoretically, according to the `Cutkosky rule', 
one expects an enhancement at $\sqrt{s}\simeq 161$ GeV $\simeq{2M_{W}}$,
the threshold for on-shell $W W$ gauge boson production, at which 
the four-point functions develop an imaginary part. 
In Fig.~\ref{fig2}(a) the ratio $R_\sigma = \sigma_{tot}/\sigma_0$
of the integrated total cross section 
$\sigma_{tot}$ to $\sigma_0$, the cross section of the low 
energy calculation of Ref.~\cite{Pham:2000bz}, is plotted for sample values of 
the Majorana masses. 
The enhancement due to the threshold singularity of the loop amplitude
is more pronounced for values of Majorana masses close to $M_W$ and 
is drastically reduced increasing $M_{N_i} \approx M_{N_j}$ to  
${\cal O}$(TeV). 
As $R_\sigma \to 1$ as $\sqrt{s} <<  M_W$ in all the cases, 
the agreement of our full calculation 
with the result of Ref.~\cite{Pham:2000bz} in the regime of 
low energies is evident.\footnote{It should be mentioned 
that the agreement, at low energies, of our Eq.(7) with Eq.(9) of 
Ref.\protect\cite{Pham:2000bz} is up to 
a factor of $4$. We have contacted the author of Ref.\protect\cite{Pham:2000bz}
on this matter and he agrees with our Eq.(7). That is, Eq.9 of 
Ref.\protect\cite{Pham:2000bz} should be  divided by $4$ 
(in Ref.\protect\cite{Pham:2000bz} the average over 
initial spins was left out~\protect\cite{private}) 
and then for energies $\sqrt{s} <<  M_W$ it 
coincides exactly with our Eq.(7).}

\begin{figure}
\begin{center}
\scalebox{0.6}{\includegraphics*[25,300][540,700]{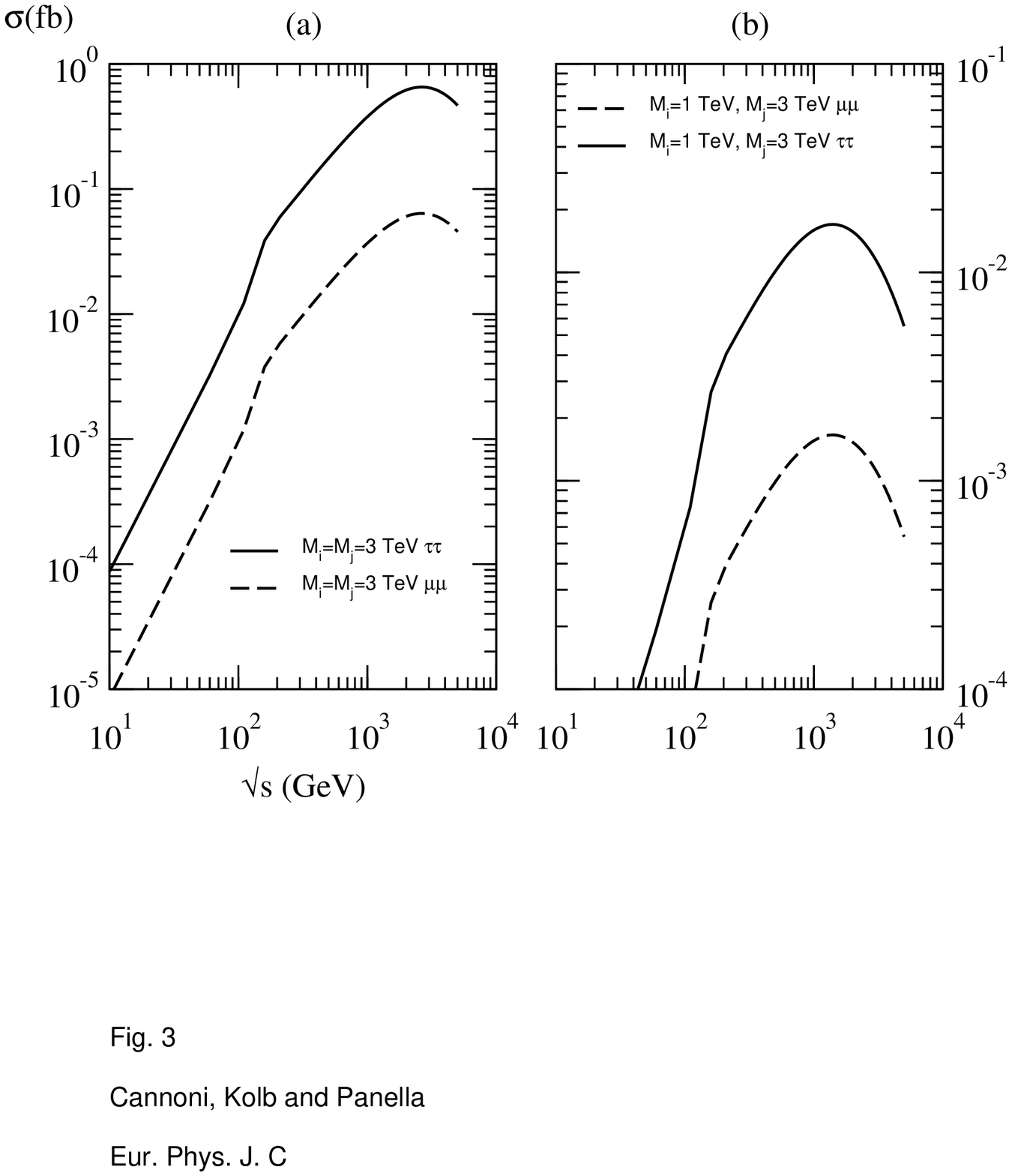}}
\caption{
Total cross sections as function of the center of mass energy, $\sqrt{s}$. 
The value of the mixing coefficients are discussed in the text.
In part (a) the solid curve referes to the case of 
$e^-e^- \to \tau^-\tau^-$ with $M_{N_i} = M_{N_j}= 3$ TeV,
while the dashed line referes to  ($ e^-e^- \to \mu\mu$) with the 
same values of Majorana masses. In (b) the Majorana masses are changed to 
somewhat lower values: $M_{N_i} = 1$ TeV, $ M_{N_j}= 3$ TeV. 
}
\label{fig3}
\end{center}
\end{figure}

The threshold effect appears to be quite spectacular 
only for values of Majorana 
masses that correspond to cross sections too small to be measured 
even at a next linear collider. 
In Fig.~\ref{fig2}(b)
the effect of the threshold singularity in the loop integral is shown 
reporting absolute cross sections for a particular choice of Majorana masses: 
$M_{N_i} = 150$ GeV and $M_{N_j} = 450$ GeV. 
The low energy 
approximation (dashed line) obtained neglecting external momenta 
in the loop is inadequate when 
the energy of the reaction increases to values comparable with the masses.
Increasing the energy, after reaching a maximum, 
the cross section starts to decrease until the asymptotic behaviour 
${\cal{O}}({1}/{s^2})$ of the loop integral $K$ is reached. 
This happens for every value of 
heavy Majorana neutrino masses and we checked numerically that, as expected,
for higher masses the asymptotic regime is reached at 
higher values of $\sqrt{s}$. 
In fact from Fig.~\ref{fig3} we note that the cross section grows with increasing 
HMN masses. 
The main contribution comes 
from the graph with two Goldstone bosons since  
their coupling is proportional to $M_{{N}_{i}}$. Moreover the chiral 
structure of the coupling selects the 
mass term in the numerator of the Majorana neutrino propagators. 
When these masses are much larger then the other quantities, 
the amplitude scales like $M^{3}_{{N}_{i}} 
M^{3}_{{N}_{j}}/M^{2}_{{N}_{i}}M^{2}_{{N}_{j}}\simeq M_{{N}_{i}}M_{{N}_{j}}$, 
i.e. is proportional to the square of the heavy masses.
This fact is  the well known non decoupling of heavy fermions 
in theories with spontaneous symmetry breaking (similarly in the SM 
the top quark gives sizable radiative corrections owing to its large 
mass and  a quadratic non decoupling). 

Heavy Majorana neutrinos naturally appear in extensions of the SM with 
right-handed neutrinos 
which generate light neutrino masses through the see-saw mechanism. 
The scale of the masses $M_{R}$ is of order $10^{9-12}$ GeV with 
very small heavy-light mixing, $U_{{\ell}j}{\sim}M_{Nj}^{-1}$, 
and the cross section will be 
suppressed by inverse powers of these  masses, recovering so 
the decoupling limit that is natural in the see-saw framework. 

An interesting scenario is the model recently proposed 
in Ref.~\cite{Ma:2000cc}, where an  attempt is made to construct 
light neutrino masses with a see-saw mechanism and no new physics beyond the 
TeV scale. This is achieved by adding a new Higgs doublet, relative to the SM, 
whose neutral component develops a naturally
small vacuum expectation value, $u\sim 1$ MeV, so that 
$m_{\nu}=m_{D}^{2}/M_{N}=f^{2}u^{2}/M_{N}\sim 1$ eV
if $M_{N}\sim 1$ TeV and $f\sim 1$, with $f$ being the Yukawa coupling. 
But the heavy-light mixing is $\sim fu/M_{N}\sim 10^{-6}$, which is
too small to have phenomenological consequences. 
Further it was
shown in Ref.~\cite{JamiAslam:20001ks} that a charged Higgs boson of this model must be 
heavier than 50 TeV,
in order to satisfy the experimental bound of the $\mu \to 3e$ decay. 
So this model does not comply with the constraints from non-observation of 
lepton flavour violation. 

More interesting from the phenomenological point of view is the case
in which HMN  have masses in the TeV range with non negligible mixing.
Mass matrices that satisfy this condition can be built, using experimental 
constraints on heavy-light  mixing (including those   
from $\beta\beta_{0\nu}$~\cite{Gluza:2002vs}).
This is achieved imposing relations among the elements of the 
neutrino mass matrix in a way that the mixing is decoupled from 
mass relations and is  bounded only by 
data~\cite{Buchmuller:1991tu},\cite{Chang:1994hz}. 
Independence of the mixing matrix from the mass relation and the consequent 
possibility of violating the  Appelquist-Carazzone 
theorem~\cite{Appelquist:tg} have led 
many authors to study HMN contributions to rare processes 
like $\mu\to e\gamma$,
$\mu\to e^{+}e^{-}e^{-}$ \cite{Cheng:1991dy,Chang:1994hz,Ilakovac:1994kj,
Ilakovac:1995km,Tommasini:1995ii,Kalyniak:1996cs,Cvetic:2002jy}.
However it was recently argued in Ref.~\cite{Gluza:2002vs} that even if such a
situation is not still ruled out by present data on neutrino oscillations, 
it requires extreme fine tuning among the elements of the Dirac mass 
matrix $m_D$ and those of $M_R$.
Keeping  this in mind, we can nonetheless explore the phenomenological 
consequences of such a scenario. 
As was done in Ref.~\cite{Panella:2001wq} we take the following experimental 
upper bounds on effective heavy-light 
mixing~\cite{Nardi:1994iv,Bergmann:1998rg,Gluza:2002vs}:
\begin{eqnarray}
s^{2}_{{\nu}_{e}}&=&\sum_{N_{i}}|U_{eN_{i}}|^{2} < 0.0027, \cr
s^{2}_{{\nu}_{\mu}}&=&\sum_{N_{i}}|U_{\mu N_{i}}|^{2} < 0.005,\cr
s^{2}_{{\nu}_{\tau}}&=&\sum_{N_{i}}|U_{\tau N_{i}}|^{2} < 0.016,
\label{mixing}
\end{eqnarray}
and allow the heavy Majorana masses to vary in the TeV range. Thus
 mixing coefficients as large as advocated in Ref.~\cite{Pham:2000bz},
$(U_{eN_{i}}U_{\mu N_{j}})^{2}\simeq 10^{-1}-10^{-2}$,
could only arise in unnatural and fine tuned models~\cite{Belanger:1995nh}. 
Note that, approximatively, the cross section goes 
like $(s^{2}_{{\nu}_{e}})^{2}(s^{2}_{{\nu}_{\mu}})^{2}$,
for real matrix elements. 
In this context, the coupling of HMN to gauge bosons and leptons is fixed to $gU_{\ell N_{i}}$,
where $g$ is the $SU(2)$ gauge coupling of the SM. 
Since the width of HMN grows with  $M^{3}_N$, at a certain value it will 
happen that $\Gamma_N > M_N$, signaling a 
breakdown of 
perturbation theory. 
The perturbative limit on $M_N$ is thereby estimated requiring 
$\Gamma_N < M_N/2$, which gives an upper bound of 
$\simeq 3$ TeV~\cite{Ilakovac:1994kj,Illana:2000ic}.

In Fig.~\ref{fig3}(a) the cross section is plotted for masses up to 
this perturbative limit, using the maximally allowed value of the mixing. 
We see that for $M_{N_i} = M_{N_j} = 3 $ TeV the signal does reach the 
level of $10^{-1},10^{-2}$ fb respectively 
for the ($\tau\tau$) and the ($\mu\mu$) signals at $\sqrt{s}= 500 $ GeV, 
which for an annual integrated luminosity of $100$ fb$^{-1}$ would 
correspond respectively to 10 and 1 event/year. At higher energies,
$\cal O$ (TeV), one could get even larger event rates (30 and 3) respectively. 
The solid curve refers to $e^-e^- \to \tau^- \tau^-$: this is largest 
because the upper limits on the mixing are 
less stringent. One can also see the onset of the asymptotic regime at 
$\sqrt{s}\approx 3 $ TeV.          
Fig.~\ref{fig3}(b) shows that the cross section quickly decreases as lower
Majorana masses are considered. 

\begin{figure}
\begin{center}
\scalebox{0.6}{\includegraphics*[55,225][480,690]{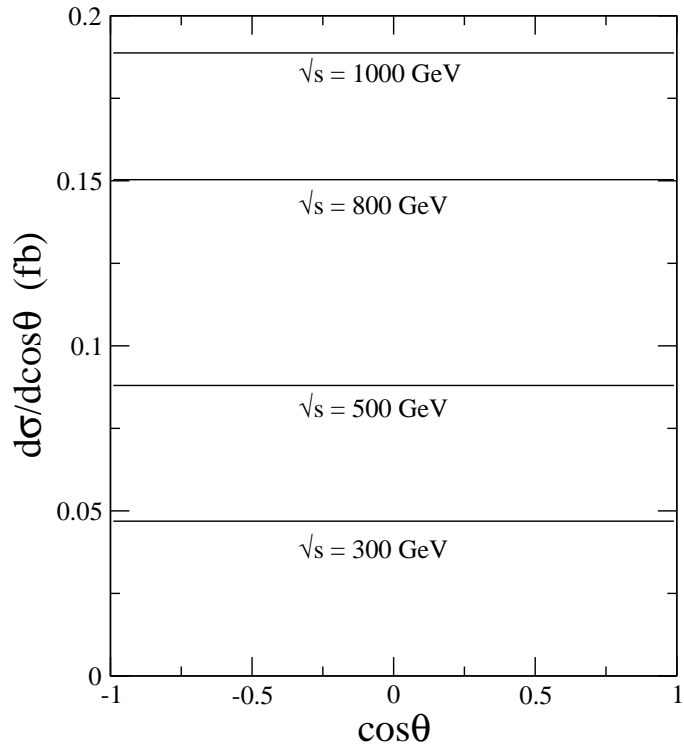}}
\caption{
Angular distribution in the polar angle of the outgoing lepton for
different values of the center of mass energy, $\sqrt{s}$ 
in the case of $e^-e^- \to \tau^-\tau^-$ with $M_{N_i} = M_{N_j}= 3$ TeV.
The curves are not exactly constant, and using an appropriate  scale 
they show small deviations from a stright line, remaining left-right
symmetric. 
}
\label{fig4}
\end{center}
\end{figure}

As even in the more optimistic cases event rates are quite modest  
it is important to check how the signal cross-section is affected by 
kinematic cuts on the angle of the outgoing leptons.  
The 
angular distributions turn out to be 
practically constant as shown in Fig.~\ref{fig4}. 
They are forward-backward symmetric because both 
the $t$ and $u$ channel are present. The absence of a strong 
dependence on the polar angle is due to the fact that within the range 
of the parameters  used here the contributing four point functions 
depend very mildly on the kinematic variables ($u$ and $t$). 
This behaviour can be most easily  understood using helicity 
amplitudes. The spinorial part common to all the diagrams is:
\begin{eqnarray}
\left[ \overline{v}(p_2)P_L u(p_1)\right]\left[\overline{u}(p_3)P_R v(p_4)
\right]=S(p_2,p_1)T(p_3,p_4),
\end{eqnarray}
that in the limit of massless external particles is a well defined 
helicity amplitude: $ e_L e_L \to \ell_L \ell_L$.
In the center of mass frame this is a S-wave scattering with $J_z=0$,
meaning that the scattered particles are emitted back to back but 
without a preferred direction relative to the collision axis ($z$). 
   
So this signal is characterized by practically flat angular distributions 
and as a result the total cross section is quite insensitive 
to angular cuts.
Values of $\sigma_T$ for different 
angular cuts are reported in Table~\ref{tab1}. With
$|\cos{\theta}|{\le}0.99$ the change  in
$\sigma_T$ is $\approx 1\% $ for all energies considered, while using 
$|\cos{\theta}|{\le}0.95$  the total cross-section decreases by 
$\approx 5\%$. Note that the reduction of the total cross section is 
measured almost precisely  by the reduction of the phase space, 
meaning that the angular distribution is constant up to $\approx 0.1\%$. 
Thus it can be concluded that 
the number of events will not be drastically affected for any 
reasonable choice of experimental cuts.  

\begin{table}
\caption{
Total cross section (for two different angular cuts at 
some sample energies). The corresponding cuts on the transverse 
momentum of outgoing leptons are also shown. 
The numerical values for masses and mixing correspond to the solid  
line of Fig. 3(a) and Fig. 4.
}
\begin{ruledtabular}
\begin{tabular}{ccccc}
&\ \  $\sqrt{s}=300$ GeV \ \ & \ \ $\sqrt{s}=500$ GeV\ \  &  
\ \ $\sqrt{s}=800$ GeV\ \  &\ \  $\sqrt{s}=1000$ GeV \ \ \cr
&$\sigma(fb)$ &$\sigma(fb)$  &$\sigma(fb)$ &$\sigma(fb)$\cr
$|\cos{\theta}|{\le}1$ & 0.094 & 0.176& 0.301 & 0.379\cr
 $|\cos{\theta}|{\le}0.99$ & 0.093 & 0.174 & 0.297 & 0.374\cr
 $|\cos{\theta}|{\le}0.95$ & 0.089 & 0.167 & 0.285 & 0.358\cr
\end{tabular} 
\end{ruledtabular} 
\label{tab1} 
\end{table}   

\begin{figure}
\begin{center}
\scalebox{0.6}{\includegraphics*[55,205][490,730]{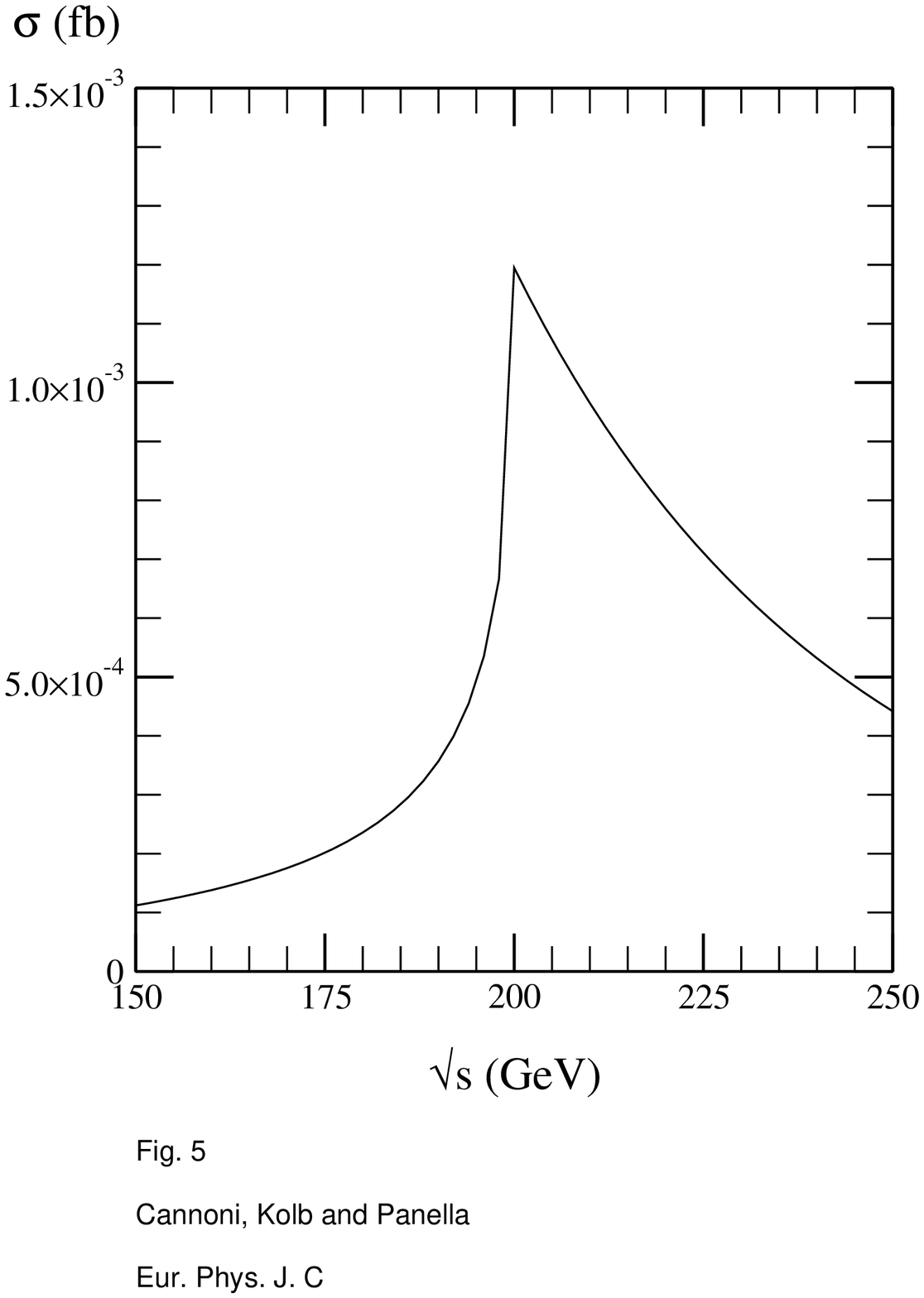}}
\caption{Total cross section for the sneutrino mediated reaction 
$\mu^-\mu^- \to \tau^-\tau^- $. 
The plot refers to the following choice of parameters:
$\Delta m_{\mu}=30$ GeV and $\Delta m_{\tau}=80$ GeV, 
$\overline{m}=m_{\chi}=100$ GeV.
}
\label{fig5}
\end{center}
\end{figure}

In supersymmetric (SUSY) extensions of the see-saw framework 
({\it e.g.}~\cite{grosshab}) the natural mass scale of the singlet 
neutrino sector is at least of order ${\cal O}(10^{12})$ GeV: in a unified
scenario such a value improves the unification of gauge coupling 
constants~\cite{Casas:2001mn}. 
Therefore HMN masses in the TeV range -although not ruled out experimentally- 
are disfavoured from a model building point of view. 
In the effective low energy SUSY see-saw framework, however, $L$ is 
violated by the light $SU(2)_L$ doublet 
sneutrinos~\cite{halprin,vergados,hirsch,grosshab}. The mass states
$\tilde{\nu}_{\ell_{1,2}}$ ($\ell$ denotes the generation) exhibit a 
mass-splitting $\Delta m_{\ell}=m_{\ell_{1}}-m_{\ell_{2}}$ that is 
constrained by its radiative contribution to neutrino 
masses~\cite{hirschconstraint}:
$\Delta m_{\ell} < 36 (156) (m_{\ell}^{exp}/1\hbox{eV})$ keV for a common 
scale of SUSY masses of 100 GeV, and 
$m^{exp}_{\ell}$ the experimental limits on neutrino masses. The two 
different values refer to average and absolute upper limits when scanning 
over the SUSY parameter space. Then, in addition to the HMN mediated 
contribution discussed above, a diagram containing $L$-violating doublet 
sneutrinos and charginos is present, see Fig.~\ref{fig1}(e). The $L$-violating 
doublet sneutrino propagator and hence the resulting cross section is 
proportional to $\Delta m$. Such a contribution has been considered
in Ref.~\cite{halprin} for the case of eV scale sneutrinos. In the realistic 
case of ${\cal O}(100)$ GeV scale sneutrinos, the exact differential 
cross section, in the notation of Fig.~\ref{fig1}(f) for the momenta, is
\begin{eqnarray} \label{sneutrinocross}
\frac{d \sigma}{d\cos\theta}&=&
\frac{1}{128\pi}\, \left(\frac{\alpha}{x_W}\right)^4 
\left|2\bigg[D_{00}(s,t)+D_{00}(s,u)\bigg] 
-\bigg[u A(s,t) + t A(s,u)\bigg]\right|^2  \, s \, \, , \\
A &=& D_2 + \sum_{i=1}^{3} D_{2i} \, . \nonumber
\end{eqnarray}
Here, the sum over the (maximally mixed) individual mass states in 
the $L$-violating sneutrino propagators is included in the loop coefficients 
$D_{00}$ and $A$ and the chargino is assumed to be a pure gaugino. In
Fig.~\ref{fig5} the resulting total $\mu^-\mu^- \rightarrow \tau^-\tau^-$
cross section is plotted for a common SUSY
mass $(m_1+m_2)/2\equiv\overline{m_{\ell}}=m_{\chi}$ = 100 GeV and for 
(maximal) values $\Delta m_{\mu}=30$ GeV and $\Delta m_{\tau}=80$ GeV 
($\Delta m_{\ell}$ must not exceed $\overline{m_{l}}$, otherwise the vacuum 
becomes 
unstable~\cite{hirsch}) allowed by the {\it kinematic} upper limits on 
neutrino masses~\cite{pdg} $m_{\mu}<$ 190 keV and $m_{\tau}<$ 18.2 MeV. Even 
for such unrealistically large neutrino masses and, correspondingly, 
unrealistic values 
for the sneutrino mass-splitting, the maximal cross section, at the
threshold singularity for real $\chi^-\chi^-$ production,  
is of order ${\cal O}(10^{-3})$ fb and therefore too small to be observable
even for a nominal integrated luminosity of 100 fb$^{-1}$/yr. 
The cross section with two colliding electrons is even smaller because
$m^{exp}_{e}<3$ eV.
Unlike the 
Majorana neutrino mediated contribution the SUSY cross section decreases 
for larger values of $\overline{m_{l}}$.
Since the reaction $e^-e^-\rightarrow \ell^- \ell^-$
conserves total lepton number, processes like $e^-e^-\rightarrow e^- \mu^-$
may arise due to interactions violating lepton flavour number 
but conserving the overall lepton number. This type of process 
will be discussed in the SUSY framework in a forthcoming paper
~\cite{inpreparation}.
         
Concluding, we have calculated the cross section for 
the process $e^{-}e^{-}\;\to\;\ell^{-}\ell^{-}$ ($\ell=\mu,\tau$) 
keeping the full dependence on the external momenta in the loop calculation 
and using the maximal value of effective light-heavy mixing angles allowed by
experiments. 
We find that only for Majorana masses in the TeV range 
the reaction has a 
measurable cross section (above $10^{-2}$ fb) with better 
prospect for the $\tau$ signal (the corresponding mixing coefficients 
being the less constrained), thereby arriving at somewhat less optimistic 
conclusions than in Ref.~\cite{Pham:2000bz}. 
We have also estimated the corresponding SUSY contribution arising
from sneutrino mixing. This, although it is affected by an 
enhancement in the region of the threshold singularity, remains below  
the minimal observable value of $10^{-2}$ fb even for the (unrealistic) 
maximal value of sneutrino mass splitting. 

\begin{acknowledgments}
The work of St. Kolb is supported by the European Union, under contract 
No. HPMF-CT-2000-00752. The authors wish to thank X~.Y.~Pham for checking
out his calculation of Ref.~\cite{Pham:2000bz}.
\end{acknowledgments}

\end{document}